\documentclass[prd,reprint,showpacs,showkeys]{revtex4-1}
\usepackage{amsfonts}
\usepackage{amssymb}
\usepackage{amsmath}
\usepackage{graphicx}
\usepackage[font={footnotesize,it}]{caption}

\begin{document}

\title{2+1-dimensional wormhole from a doublet of scalar fields}
\author{S. Habib Mazharimousavi}
\email{habib.mazhari@emu.edu.tr}
\author{M. Halilsoy}
\email{mustafa.halilsoy@emu.edu.tr}
\affiliation{Department of Physics, Eastern Mediterranean University, Gazima\~{g}usa,
Turkey. }
\date{\today }

\begin{abstract}
We present a class of exact solutions in the framework of $2+1-$dimensional
Einstein gravity coupled minimally to a doublet of scalar fields. Our
solution can be interpreted upon tuning of parameters as an asymptotically
flat wormhole as well as a particle model in $2+1-$dimensions.
\end{abstract}

\pacs{04.20.Jb; 04.60.Kz;}
\keywords{$2+1-$dimensions; Wormhole; Exact solution; Doublet scalar field}
\maketitle

\section{Introduction}

Multiplets theory of real-valued scalar fields constitutes a model that
naturally generalizes the theory of single scalar field model \cite{1}. $%
\sigma -$Model \cite{2}, the Higgs formalism \cite{3} and global monopole
theory \cite{4} are just a few to be mentioned in this category. Extra
fields amount always to extra degrees of freedom and richness in the
underlying theory. The kinetic part of the Lagrangian in this approach \ is
proportional to $\left( \nabla \phi ^{a}\right) ^{2},$ (with $a,$ a symmetry
group index) which is invariant under the symmetry transformations. In flat
spacetime this makes a linear theory but in a curved spacetime intrinsic
non-linearity automatically develops. Existence of an additional potential
is employed as instrumental to apply spontaneous symmetry breaking in the
generation of mass. Additional topological properties also are interesting
subjects in this context.

Our aim in this study firstly, is to add new degrees of freedom to scalar
fields with internal indices in the spacetime of $2+1-$dimensional gravity.
This amounts to consider multiplets of scalar fields and obtain exact
wormhole solutions in $2+1-$dimensions with non-zero curvature. $2+1-$%
dimensional wormholes were considered before \cite{5}. In this particular
dimension such a study with scalar doublets has not been conducted before.
We are motivated in this line of thought mainly by the $2+1-$dimensional
analogue of a Barriola-Vilenkin\ \cite{4}\ type global monopole solution
which is not any simpler than its $3+1-$dimensional counterpart \cite{6}. We
recall that the original idea of a spacetime wormhole, namely the
Einstein-Rosen bridge \cite{7} aimed to construct a geometrical model for an
elementary particle. For the popularity of wormholes, however, we owe to the
pioneering work of Morris and Thorne \cite{8}.

Expectedly the invariance group in our case is $O\left( 2\right) $ instead
of $O\left( 3\right) .$ It should be added that in $2+1-$dimensions even the
single scalar field solutions are very rare and restrictive \cite{9}. This
situation alone gives enough justification to search for alternatives such
as the non-isotropic scalar multiplets. Secondly, we show that the solution
obtained is a wormhole solution with the particular \ red-shift function $%
\Phi \left( r\right) =0,$ leaving us with the shape function $b\left(
r\right) $. It should be emphasized that vanishing of the red-shift function
is not a choice but rather imposed as a result of the field equations. Our
wormhole is powered by an exotic matter \cite{10} and the scalar field
doublet $\phi ^{a}\left( r,\theta \right) $ is expressed in transcendental
Lambert functions. When these are brought together our solution for the
wormhole becomes supported by a phantom scalar field doublet. Wormholes with
a phantom scalar in $3+1-$dimensions were studied in \cite{11}. Phantom
wormholes in $2+1-$dimensions were considered in \cite{12}. Another
interpretation for our solution can be considered a la' Einstein and Rosen
to represent a localized particle model in $2+1-$dimensions. We wish to
comment that $2+1-$dimensional gravity gained enough prominence during
recent decades all due to the discovery of a cosmological black hole \cite%
{13}. This gave birth to the general consensus among relativists that the $%
2+1-$dimensional geometrical structures such as black holes and wormholes
provide useful test-beds for understanding their higher dimensional cousins.
Within this context we see certain advantages in studying and understanding
better the $2+1-$dimensional wormhole solutions.

The organization of the paper is as follows. In Section II we introduce our
action and derive the field equations. We solve and plot the metric function
in Section III either as a wormhole or particle. Our conclusion in Section
IV completes the paper.

\section{Action and field equations}

The $2+1-$dimensional action in the Einstein gravity coupled to a scalar
field, without cosmological constant and self interacting potential is given
by ($16\pi G=c=1$)%
\begin{equation}
S=\int d^{3}x\sqrt{-g}\left( R-\frac{\epsilon }{2}\left( \nabla \phi
^{a}\right) ^{2}\right)
\end{equation}%
in which $\epsilon =+1/-1$ corresponds to normal / phantom scalar field
where $\phi ^{a}$ is the doublet scalar field with $a=1,2$. The standard
form of the line element for a wormhole in circularly symmetric $2+1-$%
dimensional spacetime is given by%
\begin{equation}
ds^{2}=-e^{2\Phi }dt^{2}+\frac{1}{1-\frac{b\left( r\right) }{r}}%
dr^{2}+r^{2}d\theta ^{2}.
\end{equation}%
Here $\Phi =\Phi \left( r\right) $ is the red-shift function and $b\left(
r\right) $ is the shape function satisfying the so called flare-out
conditions to which we shall refer in the sequel. Our doublet scalar field
ansatz is given by%
\begin{equation}
\phi ^{a}=\eta f\left( r\right) \frac{x^{a}}{r},
\end{equation}%
where $x^{1}=r\cos \theta $ and $x^{2}=r\sin \theta ,$ $\eta $ is a coupling
constant and $f\left( r\right) $ is a real function of $r.$ This ansatz is
well-known from the particle-like global monopole solution in the gravity
coupled field theory model \cite{4}. It admits topological properties and
due to its angular dependence it exhibits non-isotropic properties in the
radial plane. In particular the asymptotic behaviors are comparable with
those of cosmic strings which are known to possess deficit angles. Such a
model gives rise to lumpy structures in cosmic formations and naturally
modifies all tests of general relativity ranging from planetary motion to
the light bending. The reality of the model can only be tested by comparing
geodesics of all kinds with the experimental data.

Considering the doublet field given in (3) one finds 
\begin{equation}
\left( \nabla \phi ^{a}\right) ^{2}=\eta ^{2}\left( \left( 1-\frac{b}{r}%
\right) f^{\prime 2}+\frac{f^{2}}{r^{2}}\right)
\end{equation}%
such that after applying the variation of the action with respect to $f$ the
field equation becomes%
\begin{equation}
f^{\prime \prime }+\left( \Phi ^{\prime }+\frac{2r-\left( b+rb^{\prime
}\right) }{2r\left( r-b\right) }\right) f^{\prime }-\frac{f}{r\left(
r-b\right) }=0.
\end{equation}%
We note that a prime stands for the derivative with respect to $r.$
Einstein's equations are given as%
\begin{equation}
G_{\mu }^{\nu }=T_{\mu }^{\nu }
\end{equation}%
for 
\begin{equation}
T_{\mu }^{\nu }=\frac{\epsilon }{2}\left( \partial _{\mu }\phi ^{a}\partial
^{\nu }\phi ^{a}-\frac{1}{2}\partial _{\rho }\phi ^{a}\partial ^{\rho }\phi
^{a}\delta _{\mu }^{\nu }\right) .
\end{equation}%
The latter implies%
\begin{equation}
T_{t}^{t}=-\epsilon \frac{\eta ^{2}}{4}\left( \left( 1-\frac{b}{r}\right)
f^{\prime 2}+\frac{1}{r^{2}}f^{2}\right) ,
\end{equation}%
\begin{equation}
T_{r}^{r}=\epsilon \frac{\eta ^{2}}{4}\left( \left( 1-\frac{b}{r}\right)
f^{\prime 2}-\frac{1}{r^{2}}f^{2}\right)
\end{equation}%
and%
\begin{equation}
T_{\theta }^{\theta }=-T_{r}^{r}.
\end{equation}%
Accordingly, Einstein's equations read

\begin{equation}
\frac{b-rb^{\prime }}{2r^{3}}=-\epsilon \frac{\eta ^{2}}{4}\left( \left( 1-%
\frac{b}{r}\right) f^{\prime 2}+\frac{1}{r^{2}}f^{2}\right) ,
\end{equation}%
\begin{equation}
\frac{\left( r-b^{\prime }\right) \Phi ^{\prime }}{r^{2}}=\epsilon \frac{%
\eta ^{2}}{4}\left( \left( 1-\frac{b}{r}\right) f^{\prime 2}-\frac{1}{r^{2}}%
f^{2}\right)
\end{equation}%
and%
\begin{multline}
\frac{2r\left( r-b\right) \Phi ^{\prime \prime }+2\Phi ^{\prime }\left(
r\left( r-b\right) \Phi ^{\prime }+\frac{1}{2}\left( b-rb^{\prime }\right)
\right) }{2r^{2}}= \\
-\epsilon \frac{\eta ^{2}}{4}\left( \left( 1-\frac{b}{r}\right) f^{\prime 2}-%
\frac{1}{r^{2}}f^{2}\right) .
\end{multline}%
In the next section we shall find an exact solution for the four field
equations given in (5), (11), (12) and (13).

\section{Exact solutions}

The field equations admit an exact solution for $\Phi =0.$ The field
equations, in this setting become%
\begin{equation}
f^{\prime \prime }+\left( \frac{2r-\left( b+rb^{\prime }\right) }{2r\left(
r-b\right) }\right) f^{\prime }-\frac{f}{r\left( r-b\right) }=0,
\end{equation}%
\begin{equation}
\frac{b-rb^{\prime }}{2r^{3}}=-\epsilon \frac{\eta ^{2}}{4}\left( \left( 1-%
\frac{b}{r}\right) f^{\prime 2}+\frac{1}{r^{2}}f^{2}\right)
\end{equation}%
and%
\begin{equation}
\left( 1-\frac{b}{r}\right) f^{\prime 2}-\frac{1}{r^{2}}f^{2}=0.
\end{equation}%
The last equation implies%
\begin{equation}
b=\left( r-\frac{1}{r}\frac{f^{2}}{f^{\prime 2}}\right)
\end{equation}%
and upon substitution into (14) one finds that it is satisfied. Therefore
the only equation left becomes%
\begin{equation}
2rf^{\prime 2}-2ff^{\prime }-2rf^{\prime \prime }f+\epsilon \eta
^{2}r^{2}ff^{\prime 3}=0
\end{equation}%
which can be rewritten as%
\begin{equation}
\left( \frac{f}{rf^{\prime }}\right) ^{\prime }=-\frac{1}{2}\epsilon \eta
^{2}ff^{\prime }.
\end{equation}%
An integration yields%
\begin{equation}
\frac{f}{rf^{\prime }}=-\frac{\epsilon \eta ^{2}}{4}f^{2}+C_{1}
\end{equation}%
with the integration constant $C_{1}$. The resulting equation simply reads%
\begin{equation}
\frac{dr}{r}=\left( -\frac{\epsilon \eta ^{2}}{4}f+\frac{C_{1}}{f}\right) df
\end{equation}%
which is integrable as%
\begin{equation}
\ln \left( \frac{r}{r_{0}}\right) =-\frac{\epsilon \eta ^{2}}{8}%
f^{2}+C_{1}\ln f
\end{equation}%
with $r_{0}$ another integration constant. Finally, $f$ is found to be%
\begin{equation}
f=\left( \frac{r}{r_{0}}\right) ^{\xi }\exp \left[ -\frac{1}{2}LW\left( -%
\frac{\epsilon \eta ^{2}\xi }{4}\left( \frac{r}{r_{0}}\right) ^{2\xi
}\right) \right]
\end{equation}%
in which $\xi =\frac{1}{C_{1}}$ and $LW\left( x\right) $ is the Lambert-W
function \cite{14}. Using (23) we also find the exact form of $b\left(
r\right) $ which is determined as%
\begin{equation}
b=r\left[ 1-\frac{\left( 1+LW\left( -\frac{\epsilon \eta ^{2}\xi }{4}\left( 
\frac{r}{r_{0}}\right) ^{2\xi }\right) \right) ^{2}}{\xi ^{2}}\right] .
\end{equation}%
The only non-zero component of the energy momentum tensor is $%
T_{0}^{0}=-\rho $ in which the energy density is given by%
\begin{equation}
\rho =-\frac{2}{r^{2}\xi }LW\left( -\frac{\epsilon \eta ^{2}\xi }{4}\left( 
\frac{r}{r_{0}}\right) ^{2\xi }\right) .
\end{equation}%
In these solutions there are four parameters: $\eta $ and $\epsilon $ from
the action and $r_{0}$ and $\xi =\frac{1}{C_{1}}$ as integration constants.
Setting $\eta =0$ directly yields $\phi ^{a}=0$ and $b\left( r\right) =\frac{%
1}{\xi ^{2}}$ which corresponds to the flat spacetime. Due to the quadratic
form of $\eta ^{2}$, both in the action and in the solution, $\eta \lessgtr
0 $ have similar contribution. Also, $r_{0}$ is a scale factor with
dimension as $r$ and therefore we restrict $r_{0}>0.$ Unlike $\eta $, the
sign of the other two parameters bring different features for the general
solutions. Here we study each case separately.

\subsection{$\protect\epsilon =1,\protect\xi >0$}

The first setup corresponds to $\epsilon =1,\xi >0.$ In this setting, $%
f\left( r\right) $ is defined for $r<r_{c}=\left( 2/\eta \sqrt{\xi e}\right)
^{1/\xi }r_{0}$ and therefore the solution is bounded from above and we
shall call it a particle model. In this confined model the particle is
supported by normal matter with $\rho >0.$

\subsection{ $\protect\epsilon =1,\protect\xi <0$}

The second setup for the two free parameters is considered as $\epsilon =1$
and $\xi <0$. In this case the line-element can be written as%
\begin{equation}
ds^{2}=-dt^{2}+\frac{1}{B\left( r\right) }dr^{2}+r^{2}d\theta ^{2}
\end{equation}%
where 
\begin{equation}
B\left( r\right) =\frac{1}{\xi ^{2}}\left( 1+LW\left( -\frac{\eta ^{2}\xi }{4%
}\left( \frac{r}{r_{0}}\right) ^{2\xi }\right) \right) ^{2}
\end{equation}%
which is positive for $r>0.$ For $r=0$ there exists a singularity while for
large $r$, $B\left( r\right) $ asymptotes to $\frac{1}{\xi ^{2}}.$ Therefore
without loss of generality one may set $\xi =-1.$ (We note that unlike the $%
3+1-$dimensional spacetime where 
\begin{equation}
ds^{2}=-dt^{2}+\xi ^{2}dr^{2}+r^{2}\left( d\theta ^{2}+\sin ^{2}\theta d\phi
^{2}\right)
\end{equation}%
is flat only if $\xi ^{2}=1,$ in $2+1-$dimensions for any value of $\xi \neq
0$, the spacetime is flat.) In Fig. 1 we plot $B\left( r\right) $ in terms
of $r$ for various values for $\eta $. The solution is supported by normal
matter of doublet scalar field which is naked singular at $r=0$ and
asymptotically flat. We observe from this figure that larger value of $\eta
^{2}$ makes the spacetime more deviated from the flat spacetime
corresponding to $\eta ^{2}=0.$ Therefore the larger the $\eta ^{2}$ the
stronger the doublet scalar fields which results in stronger curvature.
Having $\epsilon =1$ in the action makes the scalar fields physical and also 
$\xi <0$ makes the energy density $\rho >0.$ Therefore this solution
represents a naked singular solution supported by normal doublet of scalar
field which is asymptotically flat. This solution can also be interpreted as
a particle model constructed from the doublet of scalar fields. To complete
this part, we add that the field function $f\left( r\right) $ is well
defined for $r>0$ and its asymptotic behaviors are 
\begin{equation}
\lim_{r\rightarrow 0}f\left( r\right) =\infty
\end{equation}%
and%
\begin{equation}
\lim_{r\rightarrow \infty }f\left( r\right) =0.
\end{equation}%
It is observed that the source of the field looks to be diverging at $r=0,$
where the spacetime is curved maximally and is singular.

\begin{figure}[tbp]
\includegraphics[width=70mm,scale=0.7]{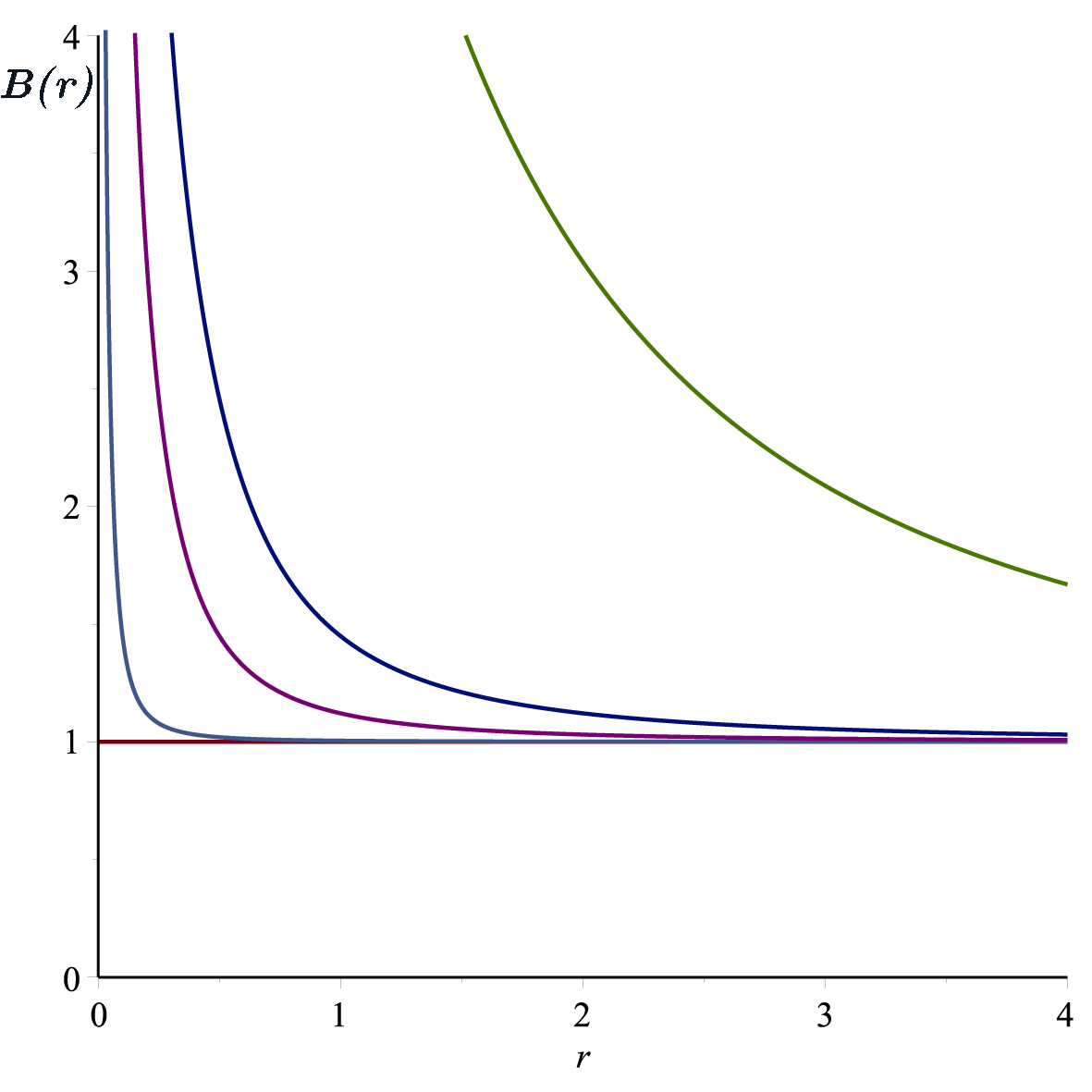} %
\captionsetup{justification=raggedright, singlelinecheck=false}
\caption{$B\left( r\right) $ versus $r$ (Eq. (27)) for various values of $%
\protect\eta =5.0,1.0,0.5,0.1$ and $0.0$ from top to bottom, respectively,
with $\protect\xi =-1,\protect\epsilon =1,r_{0}=1.$ }
\end{figure}

\begin{figure}[tbp]
\includegraphics[width=70mm,scale=0.7]{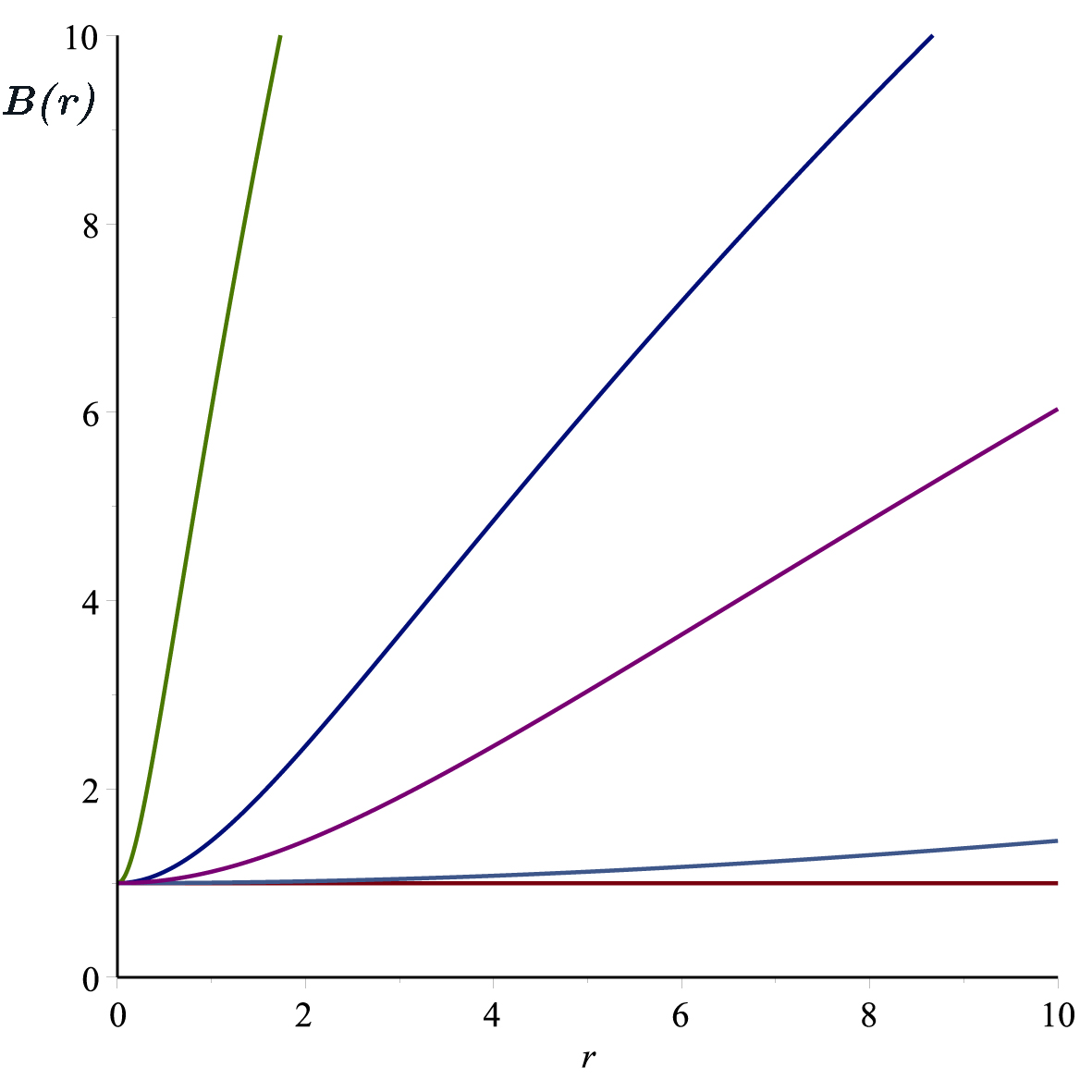} %
\captionsetup{justification=raggedright, singlelinecheck=false}
\caption{$B\left( r\right) $ versus $r$ (Eq. (27)) for various values of $%
\protect\eta =5.0,1.0,0.5,0.1$ and $0.0$ from top to bottom, respectively,
with $\protect\xi =1,\protect\epsilon =-1,r_{0}=1.$ }
\end{figure}

\begin{figure}[tbp]
\includegraphics[width=70mm,scale=0.7]{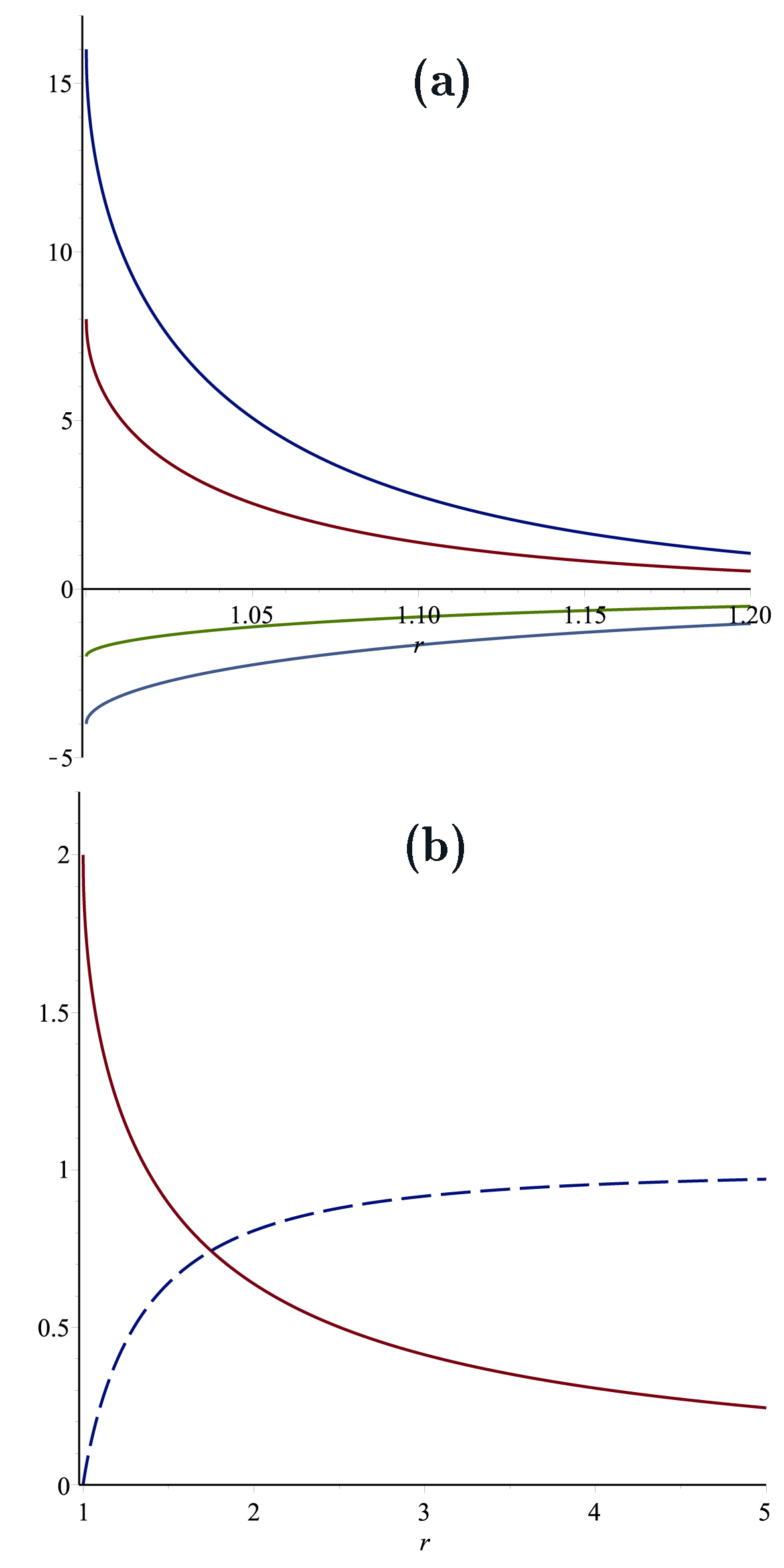} %
\captionsetup{justification=raggedright, singlelinecheck=false}
\caption{ (a) From top to bottom (Eqs. (35)-(38)); $K,$ $R_{\protect\mu 
\protect\nu }R^{\protect\mu \protect\nu },$ $\protect\rho $ and $R$ versus $%
r>b_{0}$ (b) $B\left( r\right) $ (dashed) and $\protect\eta f\left( r\right) 
$ (solid) in terms of $r$ for $r>b_{0}.$ For both we set $b_{0}=1,$ $\protect%
\epsilon =-1$ and $\protect\xi =-1.$ }
\end{figure}

\begin{figure}[tbp]
\includegraphics[width=70mm,scale=0.7]{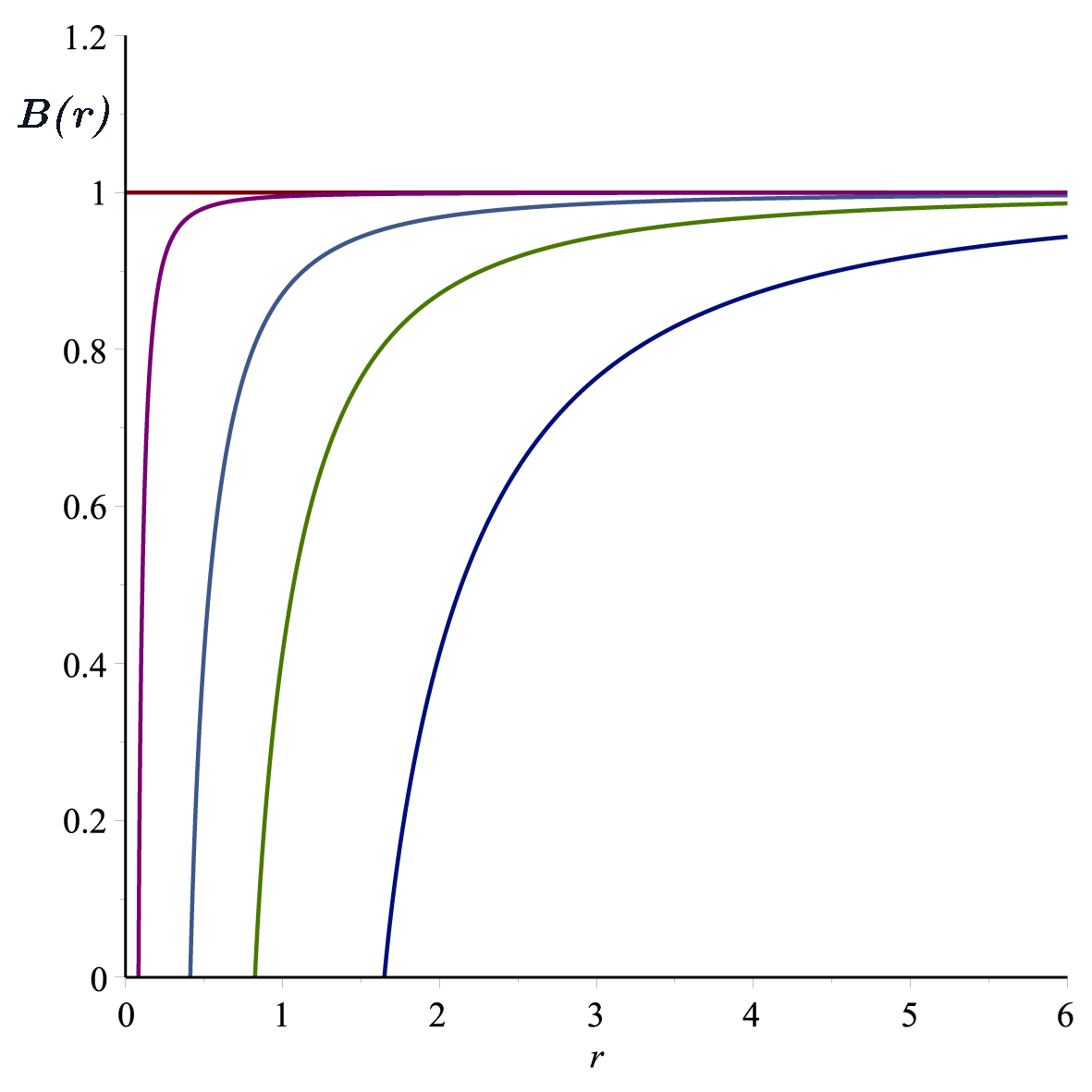} %
\captionsetup{justification=raggedright, singlelinecheck=false}
\caption{ $B\left( r\right) $ versus $r$ from Eq. (27) for various values of 
$\protect\eta =5.0,1.0,0.5,0.1$ and $0.0$ from bottom to top, respectively,
with $\protect\xi =-1,\protect\epsilon =-1,r_{0}=1.$ }
\end{figure}

\subsection{$\protect\epsilon =-1,\protect\xi >0$}

In this setting for $\epsilon =-1$ and $\xi >0$ the solution is exotic,
supported by negative energy density. The field function $f\left( r\right) $
is well defined for $r>0,$ and while at $r=0$ it vanishes at large $r$ it
diverges. \ 

In Fig. 2 we plot $B\left( r\right) $ versus $r$ for different values of $%
\eta .$ The solution is supported by exotic matter / phantom doublet of
scalar fields, which is flat near $r=0$ and non-asymptotically flat for $%
r\rightarrow \infty $. Larger value of $\eta ^{2}$ makes the spacetime more
deviated from the flat spacetime with $\eta ^{2}=0.$ Note that the
asymptotic behaviors of the solution at small $r$ and large $r$ in the
present case look to be the opposite of the previous case. The two cases are
still different solutions and by a change of variable, for instance $%
r\rightarrow \frac{1}{r},$ one can not obtain one from the other.

\subsection{ Wormhole solution for $\protect\epsilon =-1,\protect\xi <0$}

Our last general setting addresses to the most interesting case where $%
\epsilon =-1,\xi <0$ and the solution represents a wormhole with a throat
located at 
\begin{equation}
b_{0}=r_{0}\left( \frac{e\eta ^{2}\left\vert \xi \right\vert }{4}\right) ^{%
\frac{1}{2\left\vert \xi \right\vert }}
\end{equation}%
in which $e$ stands for the natural base of logarithm. The wormhole is
asymptotically flat with 
\begin{equation}
\lim_{r\rightarrow \infty }B\left( r\right) =\frac{1}{\xi ^{2}}
\end{equation}%
where we shall choose $\xi =-1.$ Both $b\left( r\right) $ and $f\left(
r\right) $ are positively defined for $r>b_{0}$ and $b\left( r\right) $
satisfies the flare-out conditions i.e., i) $b(b_{0})=b_{0}$; and ii) for $%
r>b_{0}$, $rb^{\prime }<b$ such that the field function smoothly vanishes at
infinity from its maximum value $\frac{\sqrt{2}}{\left\vert \eta \right\vert 
}$ at the throat. In terms of the throat radius one may write%
\begin{equation}
f=\frac{2\left( \frac{b_{0}}{r}\right) ^{\left\vert \xi \right\vert }}{\eta 
\sqrt{e\left\vert \xi \right\vert }}\exp \left[ -\frac{1}{2}LW\left( \frac{-1%
}{e}\left( \frac{b_{0}}{r}\right) ^{2\left\vert \xi \right\vert }\right) %
\right] ,
\end{equation}%
\begin{equation}
b=r\left( 1-\frac{1}{\xi ^{2}}\left[ 1+LW\left( \frac{-1}{e}\left( \frac{%
b_{0}}{r}\right) ^{2\left\vert \xi \right\vert }\right) \right] ^{2}\right)
\end{equation}%
with the scalar invariants given by%
\begin{equation}
K=R_{\mu \nu \alpha \beta }R^{\mu \nu \alpha \beta }=\frac{16LW\left( \frac{%
-1}{e}\left( \frac{b_{0}}{r}\right) ^{2\left\vert \xi \right\vert }\right)
^{2}}{r^{4}\xi ^{2}},
\end{equation}%
\begin{equation}
R_{\mu \nu }R^{\mu \nu }=\frac{8LW\left( \frac{-1}{e}\left( \frac{b_{0}}{r}%
\right) ^{2\left\vert \xi \right\vert }\right) ^{2}}{r^{4}\xi ^{2}}
\end{equation}%
and%
\begin{equation}
R=R_{\mu }^{\mu }=\frac{4LW\left( \frac{-1}{e}\left( \frac{b_{0}}{r}\right)
^{2\left\vert \xi \right\vert }\right) }{r^{2}\left\vert \xi \right\vert }.
\end{equation}%
The only non-zero component of the energy momentum tensor is the $tt$
component which is given by%
\begin{equation}
T_{t}^{t}=-\rho =-\frac{2LW\left( \frac{-1}{e}\left( \frac{b_{0}}{r}\right)
^{2\left\vert \xi \right\vert }\right) }{r^{2}\left\vert \xi \right\vert }.
\end{equation}%
Let's add that on the range of $r$ i.e., $r\geq b_{0}$ all of the quantities
given above are finite and asymptotically they vanish. In addition one finds%
\begin{equation}
\lim_{r\rightarrow b_{0}^{+}}f=\frac{2}{\eta \sqrt{\left\vert \xi
\right\vert }},\text{ \ }\lim_{r\rightarrow b_{0}^{+}}b=b_{0}
\end{equation}%
\begin{equation}
\lim_{r\rightarrow b_{0}^{+}}\rho =-\frac{2}{b_{0}^{2}\sqrt{\left\vert \xi
\right\vert }},\text{ \ }\lim_{r\rightarrow b_{0}^{+}}K=\frac{16}{%
b_{0}^{4}\xi ^{2}}
\end{equation}%
and%
\begin{equation}
\lim_{r\rightarrow b_{0}^{+}}R_{\mu \nu }R^{\mu \nu }=\frac{8}{b_{0}^{4}\xi
^{2}},\text{ \ }\lim_{r\rightarrow b_{0}^{+}}R=-\frac{4}{b_{0}^{2}\left\vert
\xi \right\vert }.
\end{equation}%
In Fig. 3a we plot the scalars given above to show that they are finite
everywhere and in Fig. 3b the curve of energy density $\rho $ together with
the corresponding metric function $B\left( r\right) $ are displayed. The
energy density is negative everywhere but finite, indicating the wormhole is
supported by exotic matter \cite{10}. In Fig. 4 we plot $B\left( r\right) $
versus $r$ for different values of $\eta $ with fixed values for $\xi =-1$
and $r_{0}=1$ (Note that with $r_{0}=1$ and different values for $\eta $ the
throat $b_{0}$ is not fixed). The magnitude of $\eta $ plays a critical role
to form the throat of the wormhole such that the larger value for $\eta $
implies larger size of the throat.

\subsection{$\protect\xi =0$ and $\protect\xi =\infty $}

Among the possible values for $\xi $ the case with $\xi =0$ corresponds to $%
f=1$ and consequently to the flat space solution. In contrast to that, when $%
\xi \rightarrow \infty $ the solution becomes (this can be seen from (22)
when $C_{1}=0$) 
\begin{equation}
f^{2}=-\frac{8}{\epsilon \eta ^{2}}\ln \left( \frac{r}{r_{0}}\right)
\end{equation}%
so that 
\begin{equation}
b=r\left( 1-4\left( \ln \frac{r}{r_{0}}\right) ^{2}\right)
\end{equation}%
and 
\begin{equation}
ds^{2}=-dt^{2}+\frac{dr^{2}}{\left( 2\ln \frac{r}{r_{0}}\right) ^{2}}%
+r^{2}d\theta ^{2}.
\end{equation}%
This line element has the following scalar invariants%
\begin{equation}
R=R_{\mu }^{\mu }=-\frac{8\ln \frac{r}{r_{0}}}{r^{2}}
\end{equation}%
\begin{equation}
K=R_{\mu \nu \alpha \beta }R^{\mu \nu \alpha \beta }=\frac{64\ln ^{2}\frac{r%
}{r_{0}}}{r^{4}}
\end{equation}%
and%
\begin{equation}
R_{\mu \nu }R^{\mu \nu }=\frac{32\ln ^{2}\frac{r}{r_{0}}}{r^{4}}.
\end{equation}%
It is seen clearly that $r=0$ is a spacetime singularity while at $r=r_{0}$
it is regular. This metric can't be interpreted as a wormhole since from
(43) as $r>r_{0}$ the sign of $b\left( r\right) $ turns negative which is in
contrast to the definition of a wormhole. The only non-zero component of the
energy-momentum tensor is given by%
\begin{equation}
T_{t}^{t}=\frac{4\ln \frac{r}{r_{0}}}{r^{2}},
\end{equation}%
with a divergent energy density at the origin given by%
\begin{equation}
\rho =-T_{t}^{t}
\end{equation}

\section{Conclusion}

For a number of reasons in recent decades the lower / higher dimensional
curved spacetimes received much popularity. Our aim in this paper was to
consider a doublet of non-isotropic scalar fields $\phi ^{a}\left( r,\theta
\right) $ transforming under the group $O\left( 2\right) .$ We present
parametric solutions for such a system to determine the underlying $2+1-$%
dimensional spacetime. Our solution involves the restrictive condition of
vanishing red-shift function. By making $g_{tt}=-1$ leaves us with a single
metric function $g_{rr}=\frac{1}{B\left( r\right) }$ besides $\phi
^{a}\left( r,\theta \right) .$ Once the red-shift function vanishes our
solution loses its chance to represent a black hole. However, the wormhole
and particle interpretations are admissible and as a matter of fact this
summarizes the contribution made in this paper. Our only metric function as
well as the doublet scalar functions are expressed in terms of a Lambert
function which is tabulated extensively in the literature. The source
supporting our wormhole turns out to be exotic which persists to be a
deep-rooted problem in general. We wish to remark finally that in order to
overcome this problem of exoticity we proposed recently one resolution,
which is to change the circular topological character of the throat \cite{15}%
.

\end{document}